\documentclass{article}

\usepackage{amsmath,amsthm,amsfonts}

\newcommand{\R}{\mathbb{R}}
\newcommand{\N}{\mathbb{N}}
\newcommand{\U}{\mathcal{U}}
\newcommand{\V}{\mathcal{V}}
\newcommand{\K}{\mathcal{K}}
\renewcommand{\L}{\mathcal{L}}
\renewcommand{\l}{\mathfrak{l}}
\newcommand{\cov}{\nabla}
\newcommand{\id}{\mathrm{d}}
\renewcommand{\d}{\partial}
\newcommand{\pibar}{\overline\pi}
\newcommand{\OMbar}{\overline{OM}}
\newcommand{\Mbar}{\overline{M}}
\newcommand{\fc}{\mathbf}
\newcommand{\abs}[1]{\lvert#1\rvert}
\newcommand{\absf}[1]{\lvert\fc{#1}\rvert}
\newcommand{\Abs}[1]{\left\lvert#1\right\rvert}
\newcommand{\norm}[1]{\lVert{#1}\rVert}
\newcommand{\Norm}[1]{\left\lVert{#1}\right\rVert}
\newcommand{\normf}[1]{\lVert\fc{#1}\rVert}
\newcommand{\snormf}[2]{\lVert\fc{#2}\rVert_{#1}}
\newcommand{\Phis}[1]{\Phi^s_{\scriptscriptstyle{#1}}}

\theoremstyle{plain}
\newtheorem{theorem}{Theorem}
\newtheorem{proposition}{Proposition}
\newtheorem{lemma}{Lemma}
\newtheorem{corollary}{Corollary}

\theoremstyle{definition}

\theoremstyle{remark}

\newtheorem*{note}{Note}

\begin{document}

\title{Degeneracy of the b-boundary\\ in General Relativity}
\author{Fredrik St{\aa}hl\footnote{Department of Mathematics,
University of Ume{\aa}, 901 87 Ume{\aa}, Sweden}
\footnote{E-mail: {\tt Fredrik.Stahl@math.umu.se}}}
\date{June 5, 1999}

\maketitle

\begin{abstract}
	The b-boundary construction by B.~Schmidt is a general way of 
	providing a boundary to a manifold with connection 
	\cite{Schmidt:b-boundary}.  It has been shown to have undesirable 
	topological properties however.  C.~J.~S.~Clarke gave a result 
	showing that for space-times, non-Hausdorffness is to be expected in 
	general \cite{Clarke:analysis-sing}, but the argument contains some 
	errors.  We show that under somewhat different conditions on the 
	curvature, the b-boundary will be non-Hausdorff, and illustrate the 
	degeneracy by applying the conditions to some well known exact 
	solutions of general relativity.
\end{abstract}

\section{Introduction}
\label{sec:intro}

A serious limitation in our understanding of singularities in general 
relativity is the fact that singularities by definition are not parts 
of the space-time manifold.  So in order to study the structure of 
singularities we would like to have some procedure for attaching an 
abstract boundary set containing the singular points to a space-time.  
At the very least the extended space-time should have a suitable 
topology making it possible to make statements like `close to the 
singularity' mathematically precise.

One of the candidates is the b-boundary construction by B.~Schmidt 
which works for any manifold with connection 
\cite{Schmidt:b-boundary}, and in the Lorentzian case it can be shown 
to be well-defined and locally complete 
\cite{Schmidt:local-b-completeness}.  However, for the FLRW and 
Schwarzschild space-times, the boundary is not Hausdorff separated 
from interior points \cite{Bosshard:b-boundary,Johnson:b-boundary}.  
This is a serious drawback since all points in space-time are then 
`close' to a given boundary point, making all statements about 
neighbourhoods of the singularity useless.

The b-boundary structure is closely related to the singular holonomy 
group \cite{Clarke:sing-holonomy}.  The methods used by Bosshard 
\cite{Bosshard:b-boundary} and Johnson \cite{Johnson:b-boundary} are 
heavily dependent on the specific geometry of the FLRW and 
Schwarzschild space-times, based on the study of the boundary of 
two-dimensional sections.  Clarke used a more general approach to find 
sufficient conditions for the topology to be non-Hausdorff 
\cite{Clarke:analysis-sing}.  The condition involves the asymptotic 
behaviour of the Riemann tensor and its inverse and derivative in a 
parallel propagated frame along a curve ending at the boundary point.

The argument in \cite{Clarke:analysis-sing} contains some errors 
however.  We show that under somewhat different conditions on the 
Riemann tensor and its inverse and derivative, the boundary fibres of 
the frame bundle are degenerate.  We also confirm that the conditions 
hold in the FLRW (with expansion factor $t^c$ with $c\in(0,1)$, which 
is a bit more general than in \cite{Clarke:analysis-sing}), Kasner, 
Schwarzschild, Reissner-Nordstr\"om and Tolman-Bondi space-times.

Our reasoning will depend a lot on the work by Clarke 
\cite{Clarke:analysis-sing}, the most essential difference being that 
we choose to work with small circles instead of squares in 
\S\ref{sec:ppRiemann} and that we use a stronger restriction initially 
on the derivative of the Riemann tensor.  The outline of the paper is 
as follows. In \S\ref{sec:prelim}, we introduce some notation and 
definitions.  In \S\ref{sec:ppRiemann} we approximate Lorentz 
transformations resulting from parallel propagation along small 
circles in terms of the Riemann tensor, and in \S\ref{sec:generating} 
we use these results to find a curve generating a given Lorentz 
transformation by parallel propagation.  \S\ref{sec:holon} is 
concerned with singular holonomy and gives the connection to the 
b-boundary, and we illustrate the implications for some well known 
space-times in \S\ref{sec:tot-degen}.  We also discuss some other 
contributions to the singular holonomy group in 
\S\ref{sec:part-degen}.

\section{Preliminaries}
\label{sec:prelim}

Throughout this paper, $(M,g)$ is a space-time, i.e.\ a smooth 
4-dimensional connected orientable and Hausdorff manifold $M$ with a 
smooth metric $g$ of signature $(-$$+$$+$$+)$.  The construction of 
the b-boundary may be carried out in different bundles over $M$ (see 
Refs.\ \cite{Schmidt:b-boundary}, \cite{Dodson:edge-geometry} or 
\cite{Hawking-Ellis} for some background).  Here we choose to work in 
the bundle of pseudo-orthonormal frames $OM$, consisting of all 
pseudo-orthonormal frames at all points of $M$.  $OM$ is a principal 
fibre bundle with the Lorentz group $\L$ as its structure 
group.  We write the right action of an element $\fc{L}\in\L$ as 
$R_\fc{L}:E\mapsto E\fc{L}$ for $E\in OM$.  We will, a bit sloppily, restrict 
attention to one of the connected components of $OM$ and the component 
of identity in $\L$ using the same notation.

From the fibre bundle structure of $OM$, we have a canonical 1-form 
$\theta$ which is $\R^4$-valued, and from the metric on $M$ we 
construct the connection form $\mathbf{\omega}$, which takes values in 
the Lie algebra $\l$ of $\L$ \cite{Kobayashi-Nomizu-I}.  The Schmidt 
metric on $OM$ is the Riemannian metric $G$ given by
\begin{equation}\label{eq:G-OM}
	G(X,Y) := \langle\theta(X),\theta(Y)\rangle_{\R^4}
	      	+ \langle\omega(X),\omega(Y)\rangle_{\l}
\end{equation}
where $\langle\cdot,\cdot\rangle_{\R^4}$ and 
$\langle\cdot,\cdot\rangle_{\l}$ are Euclidian inner products with 
respect to fixed bases in $\mathbb{R}^4$ and $\l$, respectively 
\cite{Schmidt:b-boundary,Dodson:edge-geometry}.

If $\kappa$ is a curve in the bundle of pseudo-orthonormal frames $OM$, we 
denote the b-length of $\kappa$ by $l(\kappa)$.  By a slight abuse 
of notation we will also write $l(\gamma,E_0)$ for the b-length, or 
generalised affine parameter length, of a curve $\gamma(t)$ in $M$, 
with respect to a given frame $E_0$ at some point on $\gamma$.  The 
definition is
\begin{equation}
	l(\gamma,E_0) := 
	\int\biggl(\sum_{i=1}^n(\fc{V}^i)^2\biggr)^{\frac12}\,\id t,
\end{equation}
where $\fc{V}^i$ are the components of the tangent vector of $\gamma$ 
with respect to the frame $E$ obtained by parallel propagation of $E_0$ 
along $\gamma$.  The notation is motivated by the fact that 
$l(\gamma,E_0)$ is the same as the b-length of the horizontal lift of 
$\gamma$ through $E_0$ in $OM$.  We also write $d(E,F)$ for the b-metric 
distance between two points $E$ and $F$ in $OM$, and $B_r(E)$ for the 
open ball in $OM$ with centre at $E$ and radius $r$.

The Schmidt metric was used by Schmidt \cite{Schmidt:b-boundary} to 
construct a boundary, the b-boundary, of the base manifold $M$, 
providing endpoints for all b-incomplete inextendible curves.  
Basically the procedure is as follows.
\begin{enumerate}
	\item Construct the Cauchy completion $\OMbar$ of $OM$ and extend 
				the group action to $\OMbar$.
	\item Let $\Mbar$ be the set of orbits of $\L$ in $\OMbar$, and 
				define a projection $\pibar:\OMbar\to\Mbar$ taking a point in 
				$\OMbar$ to the orbit through the point.
	\item $\Mbar$ is then a topological space with the topology 
				inherited from $\OMbar$ via $\pibar$, and we may identify 
				$\pibar(OM)$ with $M$.
	\item Define the b-boundary as $\d M=\Mbar\setminus M$.
\end{enumerate}
The topological space $\OMbar$ is no longer a fibre bundle since the 
action of $\L$ might be non-free on a boundary `fibre' 
(orbit).  We quantify the boundary fibre degeneracy by defining the 
singular holonomy group as
\begin{equation}
	\Phis{OM}(E) := \{\fc{L}\in\L;E\fc{L}=E\},
\end{equation}
for $E\in\pibar^{-1}(p)$ with $p\in\d M$ \cite{Clarke:sing-holonomy}.  
It follows that the boundary fibre $\pibar^{-1}(p)$ is homeomorphic to 
$\L/\Phis{OM}(E)$.  We say that the boundary fibre is degenerate 
if the singular holonomy group is nontrivial, and totally degenerate 
if the singular holonomy group is the whole Lorentz group 
$\L$.  The importance of total degeneracy is illustrated by 
the following result from \cite{Clarke:analysis-sing}.

\begin{proposition}\label{pr:degen-nonhaus}
	If $p\in\d M$ with $\pibar^{-1}(p)$ totally degenerate, then every 
	neighbourhood of $p$ in $\Mbar$ contains all null geodesics in $M$ 
	ending at $p$. In particular, $\Mbar$ is not Hausdorff.
\end{proposition}

In what follows we will need various norms, given a fixed frame $E\in 
OM$.  We use bold symbols for the array of frame components of a 
tensor in the frame $E$.  For tangent vectors $X$, we define the norm 
$\absf{X}$ to be the Euclidian norm of the frame component array 
$\fc{X}$, and similarly for cotangent vectors.  In the Lie group and 
Lie algebra, we use the Euclidian norm with respect to a fixed basis, 
and for general tensors $T$ we use the mapping norm, e.g.\
\begin{equation}\label{eq:tensornorm}
	\normf{T} := \sup_{\absf{X}=\absf{Y}=1}\abs{T_{ij}X^iY^j}
\end{equation}
for a covariant 2-tensor $T$.

\section{Parallel propagation and the Riemann tensor}
\label{sec:ppRiemann}

In this section we calculate a first approximation to the Lorentz 
transformation generated by parallel propagation around a small 
circle.  First we construct a disc with suitable properties.  Let 
$f:D_l\to M$, where
\begin{equation}
	D_l := \{ (x,y)\in\R^2; \,x^2+y^2 \le l^2 \},
\end{equation}
and put 
\begin{align}
	X(x,y) &:= f_* \frac{\d}{\d x}\Bigr|_{f(x,y)} \\
	Y(x,y) &:= f_* \frac{\d}{\d y}\Bigr|_{f(x,y)}.
\end{align}
Let $(r,\theta)$ be polar coordinates on $D_l$, i.e. $x=r\cos\theta$ 
and $y=r\sin\theta$, and put
\begin{align}
	V(x,y) &:= f_* \frac{\d}{\d r}\Bigr|_{f(x,y)} \\
	Z(x,y) &:= f_* \frac{\d}{\d\theta}\Bigr|_{f(x,y)}.
\end{align}
Then
\begin{align}
	V &= \cos\theta\,X + \sin\theta\,Y \label{eq:VXY} \\
	Z &= -r\sin\theta\,X + r\cos\theta\,Y. \label{eq:ZXY}
\end{align}
Pick a pseudo-orthonormal frame $E(0,0)$ at $p:=f(0,0)$, and define 
$E(x,y)$, where $x=r\cos\theta$ and $y=r\sin\theta$, by parallel 
propagating $E(0,0)$ along the radial curves $\rho_\theta:s\mapsto 
f(s\cos\theta,s\sin\theta)$ for each $\theta\in[0,2\pi)$.  Similarly, 
let $F(x,y)$ be defined by parallel propagating $E(r,0)$ along the 
circular curve $o_r:s\mapsto f(r\cos s,r\sin s)$ for each $r\in[0,l]$.  
Let $\fc{L}(x,y)$ be the Lorentz transformation taking $E(x,y)$ to 
$F(x,y)$, i.e.\ $F=E\fc{L}$.  From now on, bold symbols denote 
component arrays with respect to the frame $E$.

\begin{lemma}\label{la:circle}
The Lorentz transformation $\fc{L}$ is given by
\begin{equation}
	\fc{L} = \exp \:-\!\!\int_0^{\theta}\!\!\!\int_0^r\!\!  
	\fc{R(V,Z)}\,\id r\,\id\theta
\end{equation}
where $\exp$ is the exponential map $\l\to\L$.
\end{lemma}

\begin{proof}
Since $F$ is parallel along each $o_r$,
\begin{equation}
	\cov_Z F = (\cov_Z E)\fc{L} + E \cov_Z \fc{L} = 0.
\end{equation}
We may view $\fc{L}$ on each $o_r$ as a curve in the 
Lorentz group $\L$ parameterised by $\theta$. Then
\begin{equation}
	E\dot{\fc{L}}\fc{L}^{-1} = -\cov_Z E
\end{equation}
where the dot denotes differentiation with respect to $\theta$.
Now let $\lambda$ be the curve in the Lie algebra corresponding to 
$\fc{L}$ by right translation, i.e.\ $\lambda$ corresponds to the 
right-invariant vector field equal to $\dot{\fc{L}}$ at $\fc{L}$ by 
$\dot{\fc{L}}=\lambda\fc{L}$. (It might seem more natural to choose left 
translation, but then we would have to solve for $\fc{L}^{-1}$ instead.) 
Thus 
\begin{equation}
	E\lambda = -\cov_Z E.
\end{equation}
Differentiating with respect to $V$ and using that $\cov_V E=0$ and 
\begin{equation}
	\cov_V\cov_Z E = R(V,Z)E
\end{equation}
we get
\begin{equation}
	\frac{\d\lambda}{\d r} = -\fc{R(V,Z)}
\end{equation}
in the frame $E$. Integrating and solving $\dot{\fc{L}} = \lambda\fc{L}$ gives
\begin{equation}
	\fc{L} = \exp \:-\!\!\int_0^{\theta}\!\!\!\int_0^r\!\! 
	\fc{R(V,Z)}\,\id r\,\id\theta.
\end{equation}
\end{proof}

\begin{corollary}\label{co:circle}
The Lorentz transformation $\Lambda$ generated by parallel propagation 
counterclockwise around the boundary of $f(D_l)$ is given by
\begin{equation}
	\Lambda = \exp \:-\!\!\int_0^{2\pi}\!\!\!\int_0^l\!\!  
	\fc{R(V,Z)}\,\id r\,\id\theta = 
	\exp \:-\!\!\int\!\!\!\int_{D_l}\!\!\fc{R(X,Y)}\,\id\sigma
\end{equation}
where $\id\sigma$ is the area element of $D_l$ with respect to the 
metric $\id x^2 + \id y^2$.
\end{corollary}

\begin{proof}
The first expression follows immediately by letting $r \to l$ and 
$\theta \to 2\pi$ in Lemma~\ref{la:circle}.  Using \eqref{eq:VXY} and 
\eqref{eq:ZXY} and the symmetries of the Riemann tensor we get
\begin{equation}
	R(V,Z) = rR(X,Y)
\end{equation}
and hence the second formula.
\end{proof}

Let $\gamma$ be the loop at $p$ obtained by following the radial curve 
$\rho_0$, the boundary $o_l$ of the disc in the counterclockwise 
direction, and back again along $\rho_0^{-1}$.  Then parallel 
propagation along $\gamma$ generates $\Lambda$ since $E$ is parallel 
along $\rho_0$.  Suppose that $f$ is chosen such that the radial 
curves $\rho_\theta$ are geodesics and $\absf{X}=\absf{Y}=1$ at $p$, 
i.e.\ $f$ is basically the exponential map $T_p M\to M$, restricted to
\begin{displaymath}
	\{(xX_p+yY_p \in T_p M; x^2+y^2\le l^2\}.
\end{displaymath}
We can then approximate $\Lambda$ by an expression involving the value 
of $\fc R$ only at $p$.  The essential thing here is the length 
estimate.

\begin{note}
Whilst $f$ is smooth by construction, it need not be an embedding or 
even 1--1. In such a case, $E$ is not a frame field on $D_l$, but 
the construction still works. 
\end{note}

\begin{lemma}\label{la:approx}
Suppose that $\cov_V V = 0$, $\absf{X}=\absf{Y}=1$ at $p$, and $l > 0$ 
is sufficiently small for there to be an $\alpha<1$ such that
\begin{align}
	l^2\snormf{D_l}{R} &\le 10^{-3}\alpha
		\tag{I}\label{eq:approx-condR} \\
	l^3\snormf{D_l}{\cov R} &\le 10^{-6}\alpha^2
		\tag{II}\label{eq:approx-condDR}
\end{align}
where $\snormf{D_l}\cdot := \sup_{f(D_l)}\normf\cdot$. Then 
\begin{equation}
	\Norm{\Lambda - \delta + \pi l^2 \,\fc{R(X,Y)}|_p} 
	< 10^{-5}\alpha^2
\end{equation}
where $\delta$ is the identity element of $\L$, and the 
b-length of $\gamma$ is less than $9l$.
\end{lemma}

\begin{proof}
Note that $\cov_V V=0$ implies that $\absf{V}=1$ on the whole disc.  
First we need estimates for $\absf{Z}$ and $\absf{\cov_V Z}$.  Since 
$[V,Z]=0$,
\begin{equation}
	\cov^2_V Z = \cov_V\cov_Z V = R(V,Z)V,
\end{equation}
so
\begin{equation}
	\fc{\cov_V Z} = \fc{\cov_V Z}|_p + \fc{R(V,Z)V}|_{r=\xi_1}\,r
\end{equation}
and
\begin{equation}
	\fc{Z} = \fc{Z}_p + \fc{\cov_V Z}|_p\,r + 
	\frac12\fc{R(V,Z)V}|_{r=\xi_2}\,r^{2}
\end{equation}
for some $\xi_1, \xi_2 \in [0,r]$.  But from \eqref{eq:ZXY}, $\fc{Z}_p 
= 0$ and 
\begin{equation}
	\fc{\cov_V Z}|_p = -\sin\theta\,\fc{X}_p + \cos\theta\,\fc{Y}_p,
\end{equation}
so
\begin{equation}
	\absf{Z} \le r + \frac12\snormf{D_l}{R}\,\abs{\fc{Z}}_{r=\xi_2}\,r^2 \le 
	r + \frac{\alpha}{2000}\abs{\fc{Z}}_{r=\xi_2},
\end{equation}
and since $\alpha<1$,
\begin{equation}\label{eq:Zbound}
	\absf{Z} < \frac{2000}{1999}\,r,
\end{equation}
and
\begin{equation}\label{eq:DVZbound}
	\absf{\cov_V Z} < 1 + \frac{2}{1999}\,\alpha < \frac{2001}{1999}.
\end{equation}
Put
\begin{equation}
	\lambda := -\!\!\int_0^{2\pi}\!\!\!\int_0^l\!\! 
	\fc{R(V,Z)}\,\id r\,\id\theta.
\end{equation}
Then
\begin{equation}\label{eq:lambdabound}
	\norm\lambda \le 2\pi\snormf{D_l}{R}\int_0^l\absf{Z}\,\id r < 
	\frac{\alpha}{300},
\end{equation}
so
\begin{equation}\label{eq:expbound}
	\norm{\Lambda - \delta - \lambda} \le 
	\sum_{k=2}^{\infty} \frac{\norm\lambda^k}{k!} < 
	\norm\lambda^2 \sum_{k=0}^{\infty} \frac{\norm\lambda^k}{2^k} < 
	\frac{\alpha^2}{80000}.
\end{equation}

Next we replace the integral in $\lambda$ with an expression 
involving only the value of the Riemann tensor at the origin. The 
mean value theorem gives
\begin{equation}
	\fc{R(V,Z)} = \fc{R(V,Z)}|_p + 
	\fc{\cov_V\bigl(R(V,Z)\bigr)}|_{r=\xi_3}\,r
\end{equation}
for some $\xi_3 \in [0,r]$. Since $\fc{Z}_p=0$ and $\cov_V V=0$, 
\begin{equation}
	\fc{R(V,Z)} = \fc{(\cov_V R)(V,Z)}|_{r=\xi_3}\,r + 
	\fc{R(V,\cov_V Z)}|_{r=\xi_3}\,r.
\end{equation}
Applying the mean value theorem again to the first factor in the last 
term and using that $\fc{R(V,\cov_V Z)}|_p = \fc{R(X,Y)}|_p$ and 
$\cov^2_V Z = R(V,Z)V$ gives
\begin{multline}
	\fc{R(V,Z)} = \fc{(\cov_V R)(V,Z)}|_{r=\xi_3}\,r + \fc{R(X,Y)}|_p\,r \\
	+ \fc{(\cov_V R)(V,\cov_V Z)}|_{r=\xi_4}\,\xi_3\,r
	+ \fc{R\bigl(V,R(V,Z)V\bigr)}|_{r=\xi_4}\,\xi_3\,r
\end{multline}
for some $\xi_4\in[0,\xi_3]$. Thus from \eqref{eq:Zbound} and 
\eqref{eq:DVZbound},
\begin{equation}
	\Norm{\fc{R(V,Z)} - \fc{R(X,Y)}|_p\,r} < 
	\frac{4001}{1999}\snormf{D_l}{\cov R}\,r^2 + 
	\frac{2000}{1999}\snormf{D_l}{R}^2\,r^3.
\end{equation}
Integrating and using condition~\eqref{eq:approx-condR} and 
\eqref{eq:approx-condDR} along with $\alpha < 1$ we get
\begin{equation}\label{eq:pointbound}
	\Norm{\lambda + \pi l^2 \,\fc{R(X,Y)}|_p} <
	10^{-6}\alpha^2.
\end{equation}
Adding \eqref{eq:expbound} and \eqref{eq:pointbound} and 
applying Corollary~\ref{co:circle} gives
\begin{equation}
	\Norm{\Lambda - \delta + \pi l^2 \,\fc{R(X,Y)}|_p} < 
	10^{-5}\alpha^2,
\end{equation}
and we have established the first part of the lemma.  The b-length of 
$\gamma$ is given by
\begin{equation}
	l(\gamma,E) = l(\rho_0,E) + l(o_l,E) + l(\rho_0,E\Lambda).
\end{equation}
Now $\rho_0$ is a geodesic with $\abs{\dot{\rho}_0}=\absf{V}=1$, so 
the first and third terms are
\begin{equation}\label{eq:rho0len}
	l(\rho_0,E)=l
\end{equation}
and
\begin{equation}\label{eq:rho0Lambdalen}
	l(\rho_0,E\Lambda) \le l\norm\Lambda \le l \exp(\norm\lambda) < 1.1\,l
\end{equation}
by \eqref{eq:lambdabound}. The second term is
\begin{equation}
	l(o_l,E) = \int_0^{2\pi} \abs{\fc{L}^{-1}\fc{Z}}_{r=l} \,\id\theta \le 
	\int_0^{2\pi} \absf{Z}\normf{L}_{r=l} \,\id\theta
\end{equation}
since the norm of a Lorentz transformation equals the norm of its 
inverse. But $\fc{L}$ is given by Lemma~\ref{la:circle}, and applying 
\eqref{eq:Zbound}, condition~\eqref{eq:approx-condR} and $\alpha<1$ we 
get
\begin{equation}
	\normf{L}_{r=l} \le 
	\exp\Bigl( \frac{1000}{1999}l^{2}\snormf{D_l}{R}\,\theta \Bigr) < 
	\exp\Bigl( \frac{\theta}{1999} \Bigr),
\end{equation}
so using \eqref{eq:Zbound} again gives 
\begin{equation}\label{eq:ollen}
	l(o_l,E) < 
	\frac{2000}{1999}\,l 
		\int_0^{2\pi}\!\! \exp\Bigl( \frac{\theta}{1999} \Bigr) \,\id\theta
	< 6.3\,l.
\end{equation}
Adding \eqref{eq:rho0len}, \eqref{eq:rho0Lambdalen} and 
\eqref{eq:ollen} together we get the desired bound on $l(\gamma,E)$.
\end{proof}

\begin{note}
In \cite{Clarke:analysis-sing}, parallel propagation around a small 
square starting at one of the corners is investigated.  The central 
result is Lemma~2.2.1, where the conditions $l^2\normf{R}<\alpha/28$ 
and $l\normf{\cov R}<\normf{R}/20$ are used to establish
\begin{equation}
	\norm{ \Lambda - \delta - l^2\,\fc{R}(\fc{X},\fc{Y})|_p }
	< 6\alpha^2.
\end{equation}
An explicit calculation in FLRW space-time illustrates that this is 
impossible without using a stronger condition on $\normf{\cov R}$.  
Apart from typographical errors, the main problem seems to be in the 
argument at the top of page~24 of \cite{Clarke:analysis-sing}.  It is 
possible to obtain an estimate of order $\alpha^2$ for parallel 
propagation around a circle with the starting point at the centre with 
a bound of order $\alpha$ on $\normf{\cov R}$, by modifying the 
argument in our Lemma~\ref{la:approx}.  The idea is to use a second 
order expansion of $\fc{R(V,Z)}$ and then a symmetry argument to get 
rid of the first $\cov\fc{R}$ term.  However, the penalty for the 
weaker condition on $\cov\fc{R}$ is that a condition on 
$\norm{\cov^2\fc{R}}$ of order $\alpha^2$ has to be imposed.  For our 
purpose, condition \eqref{eq:approx-condDR} is sufficient.
\end{note}

\section{Generating Lorentz transformations}
\label{sec:generating}

Using the approximation from Lemma~\ref{la:approx}, we can construct 
a loop generating a given Lorentz transformation exactly, provided 
that the transformation is sufficiently close to the identity.  The 
idea is to generate a sequence of approximate transformations by 
parallel propagation along the boundaries of a sequence of 
appropriately constructed circles, applying Lemma~\ref{la:approx} 
at each stage.  First we construct the approximate curves to be used 
as building blocks for the final curve.

\begin{note}
To ensure the existence of the disks used to generate the curves, we 
need to avoid the situation where one of the radial curves cannot be 
continued because it runs into a singularity. If we restrict attention 
to a subset $\U$ of $OM$ with compact closure, this can only happen 
if $\U$ contains a trapped inextendible incomplete curve 
\cite{Schmidt:local-b-completeness}. This is avoided if we assume 
that the closure of $\U$ in $\OMbar$ is compact and contained in $OM$.
\end{note}

\begin{lemma}\label{la:gener1}
Let $\lambda\in\l$, $E\in OM$ and $p=\pi(E)$ be given and suppose that 
there is a bivector $\fc{W}$ such that $\fc{R}_p(\fc{W})=\lambda$, 
where $\fc{R}_p$ is the Riemann tensor in the frame $E$ at $p$.  Put
\begin{equation}
	\U := \{F;d(E,F) < 22\normf{W}^{1/2}\}
\end{equation}
and $\snormf{\U}\cdot:=\sup_{\U}{\norm\cdot}$, and let 
$\fc{L}:=\exp\lambda$.  Also, assume that $\normf{W}$ is sufficiently 
small for the closure of $\U$ in $\OMbar$ to be compact and contained 
in $OM$.  If
\begin{align}
	\normf{W} &< (\pi/4000) \,\snormf{\U}{R}^{-1}
		\tag{I}\label{eq:gener1-condR}  \\
	\normf{W} &< (\pi/40000) \,\snormf{\U}{\cov R}^{-2/3}
		\tag{II}\label{eq:gener1-condDR}
\end{align}
then there is a horizontal curve $\gamma$ in $\U$ starting at $E$ 
which generates a Lorentz transformation $\Lambda$ with
\begin{equation}
	\norm{\fc{L}-\Lambda} < 10^{-3}\alpha^2,
\end{equation}
where
\begin{equation}
	\alpha < \max\left\{ 
		\frac{4000}{\pi} \normf{W} \snormf{\U}{R}, 
		\Bigl( \frac{40000}{\pi}\normf{W} \Bigr)^{3/4} \snormf{\U}{\cov R}^{1/2} 
	\right\},
\end{equation}
and the b-length of $\gamma$ is less than $22\normf{W}^{1/2}$.
\end{lemma}

\begin{proof}
We start by decomposing $\fc{W}$ as 
\begin{equation}
	\fc{W} = \fc{A}\cos\theta + \fc{*A}\sin\theta
\end{equation}
where $\fc{A}$ and $\fc{*A}$ are dual independent simple bivectors. Inverting 
this relation and using that for any bivector $\fc{B}$, 
\begin{equation}
	\normf{*B} \le 2\sqrt{3}\normf{B},
\end{equation}
we get 
\begin{equation}\label{eq:Abound}
	\normf{A},\normf{*A} < 4\normf{W}.
\end{equation}
Define a disc by $f:D_{l_1}\to\pi(\U)$, such that $\absf{X}=\absf{Y}=1$, 
$\langle\fc{X},\fc{Y}\rangle_{\R^4}=0$ and $\pi 
l_1^2\,\fc{X}\wedge\fc{Y}=-\fc{A}\cos\theta$ at $E$, as in 
\S\ref{sec:ppRiemann}.  Then \eqref{eq:Abound} gives
\begin{equation}\label{eq:lbound}
	l_1^2 < \frac{4}{\pi}\normf{W},
\end{equation}
so $f(D_{l_1})\subset\pi(\U)$. Put
\begin{equation}
	\alpha := \max\left\{ 
		10^3 \,l_1^2 \snormf{\U}{R}, 
		10^3 \,l_1^{3/2} \snormf{\U}{\cov R}^{1/2} 
	\right\}.
\end{equation}
Then condition~\eqref{eq:gener1-condR} and \eqref{eq:gener1-condDR} give 
$\alpha<1$ so Lemma~\ref{la:approx} applies.

From Lemma~\ref{la:approx} we have a loop $\gamma_1$ at $p$ and a 
Lorentz transformation $\Lambda_1$ generated by parallel propagation 
around $\gamma_1$.  Replacing $A\cos\theta$ and $l_1$ with 
$*A\sin\theta$ and $l_2$ and repeating the above procedure we get 
another loop $\gamma_2$ at $p$ which generates a Lorentz 
transformation $\Lambda_2$.  Put
\begin{equation}
	\fc{Z}_1 := \Lambda_1 - \delta - \fc{R}_p(\fc{A}\cos\theta)
\end{equation}
and
\begin{equation}
	\fc{Z}_2 := \Lambda_2 - \delta - \fc{R}_p(\fc{*A}\sin\theta).
\end{equation}
From Lemma~\ref{la:approx} we know that 
\begin{equation}\label{eq:Z1Z2bound}
	\norm{\fc{Z}_1},\norm{\fc{Z}_2} < 10^{-5}\alpha^2.
\end{equation}
Let $\Lambda=\Lambda_1\Lambda_2$.  Then $\Lambda$ is generated by 
parallel propagation around the concatenation $\gamma$ of $\gamma_1$ 
and $\gamma_2$, and we may write
\begin{equation}\label{eq:Lambda-L}
\begin{split}
	\Lambda-\fc{L}
	&= \fc{Z}_1 \bigl( \fc{Z}_2 + \delta + \fc{R}_p(\fc{*A}\sin\theta) \bigr) 
	+ \bigl( \delta + \fc{R}_p(\fc{A}\cos\theta) \bigr) \fc{Z}_2 \\ 
	&+ \fc{R}_p(\fc{A}\cos\theta)\fc{R}_p(\fc{*A}\sin\theta) - 
		\sum_{k=2}^\infty \frac{\lambda^k}{k!}.
\end{split}
\end{equation}
Using first \eqref{eq:Z1Z2bound} and \eqref{eq:Abound} and then 
condition~\eqref{eq:gener1-condR} we get
\begin{equation}\label{eq:Lambda1bound}
	\norm{\Lambda_1} 
	= \norm{ \fc{Z}_1 + \delta + \fc{R}_p(\fc{A}\cos\theta) } 
	< 10^{-5}\alpha^2 + 1 + 4\norm{\fc{R}_p}\normf{W} 
	< 1.01
\end{equation}
and similarly
\begin{equation}\label{eq:Lambda2bound}
	\norm{\Lambda_2} = 
	\norm{ \fc{Z}_2 + \delta + \fc{R}_p(\fc{*A}\sin\theta) } 
	< 1.01
\end{equation}
and
\begin{equation}\label{eq:Lambda-Zbound}
	\norm{ \delta + \fc{R}_p(\fc{A}\cos\theta) } < 1.01.
\end{equation}
Inserting \eqref{eq:Lambda2bound} and \eqref{eq:Lambda-Zbound} into 
\eqref{eq:Lambda-L} and using that 
\begin{equation}
	\norm\lambda \le \norm{\fc{R}_p}\normf{W} < \frac{\pi}{4000}
\end{equation}
and
\begin{equation}
	\normf{W}<\pi(l_1^2+l_2^2)
\end{equation}
and condition~\eqref{eq:gener1-condR} gives
\begin{equation}\label{eq:Lambda-Lbound}
	\norm{\Lambda-\fc{L}} < 
	2.02\cdot10^{-5}\alpha^2 + 
		16\norm{\fc{R}_p}^2\normf{W}^2 + 
		\frac{\norm\lambda^2}{2(1-\norm\lambda)} < 
	10^{-3}\alpha^2.
\end{equation}

The length of $\gamma$ is the sum of the lengths of $\gamma_1$ 
and $\gamma_2$. From Lemma~\ref{la:approx} and 
\eqref{eq:lbound} we find that 
\begin{equation}
	l(\gamma_1,E) < 9l_1 < 11\normf{W}^{1/2}.
\end{equation}
The same holds for $\gamma_2$ except that we have to correct for 
the starting frame being $E\Lambda_1$ instead of $E$. From 
\eqref{eq:Lambda1bound}, 
\begin{equation}
	l(\gamma_2,E\Lambda_1) < 9l_2\norm{\Lambda_1} < 11\normf{W}^{1/2},
\end{equation}
and thus
\begin{equation}
	l(\gamma,E) < 22\normf{W}^{1/2}.
\end{equation}
\end{proof}

The Riemann tensor in a given frame can be viewed as a map from the 
space of bivectors to the Lie algebra $\l$.  We use the norm
\begin{equation}
	\normf{W} := 2\sup_{\absf{X}=\absf{Y}=1}\abs{W^{ij}X_{i}Y_{j}}
\end{equation}
for the bivectors, so that the mapping norm
\begin{equation}
	\normf{R} := \sup_{\normf{W}=1}\normf{R(W)}
\end{equation}
agrees with the previously defined tensor norm \eqref{eq:tensornorm}.  
We now concentrate on the case when the Riemann tensor in the frame 
$E$ is invertible (the frame is of course not essential here since 
invertibility in one frame is equivalent to invertibility in any 
frame).  Note that if the Riemann tensor is invertible at a point 
$F\in OM$, the image of the space of bivectors is the whole Lorentz 
group $\L$, so by the standard holonomy theory the infinitesimal 
holonomy group is the whole of $\L$.  Thus the length estimate is the 
important result here.  The idea is to piece the curves from 
Lemma~\ref{la:gener1} together to generate a sufficiently small 
Lorentz transformation exactly.

\begin{lemma}\label{la:gener}
Let $\lambda\in\l$, $E\in OM$ and $p:=\pi(E)$ be given and suppose that 
$\fc{R}_p$, the Riemann tensor in the frame $E$ at $p$, is invertible.  
Let $\fc{R}_p^{-1}$ be the inverse and put
\begin{equation}
	\U := \{F;d(E,F) < 24\norm{\fc{R}_p^{-1}}^{1/2}\norm\lambda^{1/2}\}.
\end{equation}
and $\snormf{\U}\cdot := \sup_{\U}{\norm\cdot}$.  If the closure of 
$\U$ in $\OMbar$ is compact and contained in $OM$ and
\begin{align}
	\norm\lambda &< 
		10^{-6} \,\norm{\fc{R}_p^{-1}}^{-2} \snormf{\U}{R}^{-2}
		\tag{I}\label{eq:gener-condR}  \\
	\norm\lambda &< 
		10^{-12} \,\norm{\fc{R}_p^{-1}}^{-3} \snormf{\U}{\cov R}^{-2}
		\tag{II}\label{eq:gener-condDR}
\end{align}
then there is a horizontal curve in $\U$ starting from $E$ 
and ending at $E\exp\lambda$, piecewise smooth except possibly at the 
endpoint, of b-length less than
\begin{equation}
	24\norm{\fc{R}_p^{-1}}^{1/2}\norm{\lambda}^{1/2}.
\end{equation}
\end{lemma}

\begin{proof}
Let $\fc{L}:=\exp\lambda$.  To construct the first square, put 
$\fc{W}:=\fc{R}_p^{-1}(\lambda)$.  Since 
\begin{equation}
	\norm{\fc{R}_p^{-1}} \snormf{\U}{R} \ge
	\norm{\fc{R}_p^{-1}} \norm{\fc{R}_p} \ge 1,
\end{equation}
condition~\eqref{eq:gener-condR} gives
\begin{equation}\label{eq:gener-condC}
	\norm\lambda < 10^{-6}.
\end{equation}
Applying condition~\eqref{eq:gener-condR} to the first factor of 
$\norm\lambda^2$ and \eqref{eq:gener-condC} to the second factor and 
then taking the square root gives that 
condition~\eqref{eq:gener1-condR} of Lemma~\ref{la:gener1} is 
fulfilled.  Similarly, applying condition~\eqref{eq:gener-condDR} to 
the first factor of $\norm\lambda^3$, \eqref{eq:gener-condC} to the 
other two factors and taking the third root gives that 
condition~\eqref{eq:gener1-condDR} of Lemma~\ref{la:gener1} is 
fulfilled.  Thus Lemma~\ref{la:gener1} applies and we have a loop 
$\gamma_1$ which generates a first approximation $\fc{L}_1$ to 
$\fc{L}$.  Also,
\begin{equation}\label{eq:alphabound}
\begin{split}
	\alpha^2 < 
		\norm\lambda
		\max\biggl\{ 
			&\Bigl( \frac{4000}{\pi} \Bigr)^2 
				\norm\lambda \norm{\fc{R}_p^{-1}}^2 \snormf{\U}{R}^2, \\
			&\Bigl( \frac{40000}{\pi} \Bigr)^{3/2} 
				\bigl( \norm\lambda \norm{\fc{R}_p^{-1}}^3 
							 \snormf{\U}{\cov R}^2 \bigr)^{1/2}
			\biggr\}
\end{split}
\end{equation}
so from condition~\eqref{eq:gener-condR} and \eqref{eq:gener-condDR},
\begin{equation}
	\alpha^2 < 2\norm\lambda,
\end{equation}
and Lemma~\ref{la:gener1} gives
\begin{equation}\label{eq:L-L1bound}
	\norm{\fc{L}-\fc{L}_1} < \frac{1}{500}\norm\lambda.
\end{equation}
Next we repeat the construction for the Lorentz 
transformation $\fc{L}_1^{-1}\fc{L}$.  We first have to check that 
the conditions are satisfied.  But from \eqref{eq:Lambda1bound} and 
\eqref{eq:Lambda2bound},
\begin{equation}\label{eq:L1bound}
	\norm{\fc{L}_1} \le \norm{\Lambda_1} \norm{\Lambda_2} < 1.1,
\end{equation}
and from \eqref{eq:L-L1bound} and the fact that the norm of a Lorentz 
transformation equals the norm of its inverse,
\begin{equation}
	\norm{\fc{L}_1^{-1}\fc{L}-\delta} < 
	\norm{\fc{L}_1} \norm{\fc{L}-\fc{L}_1} < 
	\frac{1}{450}\norm\lambda.
\end{equation}
It follows that we can write $\fc{L}_1^{-1}\fc{L} = \exp\lambda_2$ with
\begin{equation}\label{eq:lambda2bound}
	\norm{\lambda_2} 
	< \frac{450}{449}\norm{\fc{L}_1^{-1}\fc{L}-\delta} 
	< \frac{1}{449}\norm\lambda.
\end{equation}
Thus $\lambda_2$ satisfies the conditions as long as the 
generating curve stays in $\U$.

Repeating the above process we get a series of loops $\gamma_k$ 
corresponding to a sequence $\lambda_k$ of Lie algebra elements, 
generating Lorentz transformations $\fc{L}_k$.  The products 
$\hat\fc{L}_k=\fc{L}_1\fc{L}_2\dots\fc{L}_k$ are generated by parallel 
propagation along the concatenation of the curves 
$\gamma_1,\gamma_2,\dots,\gamma_k$, and 
\begin{equation}
	\norm{\hat\fc{L}_k - \fc{L}} \le 
	\norm{\hat\fc{L}_{k-1}}\norm{\fc{L}_k - \hat\fc{L}_{k-1}^{-1}\fc{L}} < 
	1.1^{k-1} \frac{1}{500} \norm{\lambda_k}
\end{equation}
from \eqref{eq:L-L1bound} and repeated application of 
\eqref{eq:L1bound}. But \eqref{eq:lambda2bound} gives
\begin{equation}\label{eq:lambdakbound}
	\norm{\lambda_k} < \Bigl(\frac{1}{449}\Bigr)^{k-1}\norm\lambda,
\end{equation}
so $\hat\fc{L}_k \to \fc{L}$ as $k \to \infty$.  It remains to show 
that the resulting curve is contained in $\U$.  From 
Lemma~\ref{la:gener1},
\begin{equation}
	l(\gamma_1,E) < 22\norm{\fc{R}_p^{-1}}^{1/2}\norm{\lambda}^{1/2}.
\end{equation}
For $\gamma_k$, we have to take into account that the starting 
point is $E\hat\fc{L}_{k-1}$ instead of $E$, so
\begin{equation}
\begin{split}
	l(\gamma_k,E\fc{L}_{k-1}) 
	&< 22\norm{\fc{R}_p^{-1}}^{1/2}\norm{\lambda_k}^{1/2}\norm{\hat\fc{L}_{k-1}}\\
	&< 22\norm{\fc{R}_p^{-1}}^{1/2}
		 \Bigl(\frac{1}{449}\Bigr)^{(k-1)/2}\norm{\lambda}^{1/2}\,1.1^{k-1}
\end{split}
\end{equation}
from \eqref{eq:L1bound} and \eqref{eq:lambdakbound}.  Summing over $k$ 
we get the desired bound on the length, and it is evident that the 
generating curve stays in $\U$.
\end{proof}

\begin{note}
The main difference between our Lemma~\ref{la:gener} and Lemma~2.2.2 
of \cite{Clarke:analysis-sing} is that condition~\eqref{eq:gener-condR} 
involves the second power of $\fc{R}_p^{-1}$ and $\fc{R}$ instead of 
the first.  This is needed to establish \eqref{eq:alphabound} which is 
essential for the construction of the sequence of circles to work.  
The corresponding equation at the bottom of page~26 in 
\cite{Clarke:analysis-sing} is incorrect since there a bound on 
$\Gamma^2\norm\lambda$ is needed, but the given conditions only 
provide a bound on $\Gamma\norm\lambda$.
\end{note}

It is now a simple matter to generate arbitrary transformations by 
splitting them in a finite number of factors, sufficiently small for 
Lemma \ref{la:gener} to apply, and joining together the resulting 
curves.  Note that we do not need to go through the approximation 
scheme in Lemma~\ref{la:gener} more than once as is done in 
\cite{Clarke:analysis-sing}, since once we have a curve generating the 
first factor, we can translate it along the fibres to get curves 
generating the other factors.

\begin{theorem}\label{th:gener}
Let $E\in OM$ with $p:=\pi(E)$ and put 
\begin{equation}
	\U := \{F\in OM;d(E,F)<\delta\}
\end{equation}
for some $\delta>0$, small enough for the closure of $\U$ in $\OMbar$ 
to be compact and contained in $OM$.  Let $\fc{L}:=\exp\lambda$ be a 
Lorentz transformation and suppose that $\fc{R}$ is invertible on 
$\U$.  Then there is a horizontal curve $\gamma$ in 
$\pi^{-1}\circ\pi(\U)$ which generates $\fc{L}$ with
\begin{equation}
	l(\gamma,E) < 
	24 \normf{L} \norm{\fc{R}_p^{-1}}^{1/2} \norm\lambda^{1/2} n^{1/2},
\end{equation}
where 
\begin{equation}\label{eq:ndef}
	n := \left\lceil \norm\lambda \max\left\{
	10^6 \norm{\fc{R}_p^{-1}}^2 \snormf{\U}{R}^2,
	10^{12} \norm{\fc{R}_p^{-1}}^3 \snormf{\U}{\cov R}^2,
	24^2 \norm{\fc{R}_p^{-1}}/\delta^2
	\right\} \right\rceil.
\end{equation}
\end{theorem}

\begin{proof}
We start by generating the Lorentz transformation 
$\fc{L}_1:=\exp(\lambda/n)$ where $n\in\N$ is chosen sufficiently 
large for Lemma~\ref{la:gener} to hold on a subset of $\U$, which 
gives \eqref{eq:ndef}.  By Lemma~\ref{la:gener} there exists a 
horizontal curve $\gamma_1$ in $\U$ from $E$ to $E_1 := E\fc{L}_1$.  
Let $\fc{L}_k := (\fc{L}_1)^k$ and $E_k := E\fc{L}_k$ for 
$k=2,3,\dots,n$.  Then $\gamma_k = \gamma_1 \fc{L}_{k-1}$ is a 
horizontal curve from $E_{k-1}$ to $E_k$ since the action of the 
Lorentz group preserves horizontal curves.

Let $\gamma$ be the combined curve obtained by joining the 
curves $\gamma_k$ in sequence.  Then $\gamma$ generates 
$\fc{L}_n = \fc{L}$ and since 
\begin{equation}
	l(\gamma_k) \le \norm{\fc{L}_k}\,l(\gamma_1) \le \normf{L}\,l(\gamma_1)
\end{equation}
and
\begin{equation}
	l(\gamma_1) < 24\norm{\fc{R}_p^{-1}}^{1/2}\norm{\lambda/n}^{1/2},
\end{equation}
the result follows.
\end{proof}

\section{The singular holonomy group}
\label{sec:holon}

We can now relate the structure of the singular holonomy group with 
the asymptotic behaviour of the Riemann tensor.  First we need the 
following characterisation from \cite{Clarke:analysis-sing}.
\begin{proposition}\label{pr:sing-hom-char}
Suppose that $\gamma:(0,1]\to OM$ is a horizontal curve with 
$\gamma(0) = E \in \pibar^{-1}(p)$ and $p\in\d M$.  Then 
$\fc{L}\in\Phis{OM}(E)$ if and only if there is a sequence $t_i$ with 
$t_i\to 1$ and loops $\kappa_i:[0,1]\to M$ such 
that
\begin{gather}
	\kappa_i(0) = \kappa_i(1) = \pi\circ\gamma(t_i),
	\tag{I}\label{eq:sing-hom1} \\
	\fc{L}_i \to \fc{L}
	\tag{II}\label{eq:sing-hom2} \\
	l(\kappa_i,\gamma(t_i)) \to 0
	\tag{III}\label{eq:sing-hom3}
\end{gather}
where $\fc{L}_i$ are the Lorentz transformations obtained by parallel 
propagating $\gamma(t_i)$ around $\kappa_i$ for each $i$.
\end{proposition}

We may use Proposition~\ref{pr:sing-hom-char} to give an alternative 
definition of the singular holonomy group \cite{Clarke:sing-holonomy}.  
Let $\varphi_a(F)$ be the group of Lorentz transformations generated 
by parallel transport around loops $\kappa$ at $\pi(F)$ with 
$l(\kappa,F)\le a$.  Then if $\gamma:(0,1]\to OM$ is a horizontal 
curve starting at $\gamma(0)=E\in\pibar^{-1}(p)$ with $p\in\d M$,
\begin{equation}
	\Phis{OM}(E) :=
	\bigcap_{a\in\R^+}\,\overline{\bigcup_{t\in(0,1]}\varphi_a(\gamma(t))}.
\end{equation}

A nontrivial $\Phis{OM}$ may have several causes.  For example, the 
bounded part of the curvature may contribute as well as the unbounded 
part \cite{Clarke:sing-holonomy}, and non-trivial topologies can 
generate discrete subgroups (see \S\ref{sec:part-degen} below).  
In the following section we concentrate on using Lemma~\ref{la:gener} 
to show how divergence of the Riemann tensor can cause total 
degeneracy.

\section{Total Degeneracy}
\label{sec:tot-degen}

Combining Proposition~\ref{pr:sing-hom-char} with Theorem 
\ref{th:gener} we get the following sufficient conditions for 
total degeneracy of a boundary fibre.  In the rest of this section we 
will see that the conditions are indeed fulfilled in many 
interesting cases relevant to general relativity.

\begin{theorem}\label{th:degeneracy}
Suppose that $\gamma:(0,1]\to OM$ is a horizontal curve with 
$\gamma(0) = E \in \pibar^{-1}(p)$ and $p\in\d M$, and that there 
are sequences $t_i\to 0$ and $\rho_i\to 0$ such that $\fc{R}$ is 
invertible on the balls $\U_i := B_{\rho_i}(\gamma(t_i))$.  If the 
closure of each $\U_i$ in $\OMbar$ is compact and contained in $OM$ 
and $\norm{\fc{R}_i^{-1}}^3 \snormf{\U_i}{R}^2$, 
$\norm{\fc{R}_i^{-1}}^2 \snormf{\U_i}{\cov R}$ and 
$\norm{\fc{R}_i^{-1}}/\rho_i$ tend to $0$ as $t_i \to 0$, then 
$\Phis{OM}(E)=\L$.
\end{theorem}

Note that invertibility of the Riemann tensor means that it is 
injective, i.e.\ there are no 2-planes on which $\fc{R}$ vanishes, and 
surjective, i.e.\ there is no subspace of the Lie algebra unaffected 
by curvature.  If $\fc{R}$ is invertible, 
\begin{equation}
	\norm{\fc{R}^{-1}} 
	= \sup_{\lambda} \frac {\norm{\fc{R}^{-1}(\lambda)}} {\norm\lambda} 
	= \sup_{\fc{W}} \frac {\normf{W}} {\normf{R(W)}} 
	= \Bigl( \inf_{\normf{W}=1} \normf{R(W)} \Bigr)^{-1},
\end{equation}
so $\norm{\fc{R}^{-1}} \to 0$ if and only if $\normf{R(W)}$ diverges for 
all bivectors $\fc{W}$. In other words, $\norm{\fc{R}^{-1}}\to 0$ if and only 
if, for all index pairs $k$ and $l$, there are two indices $i$ and $j$ 
such that the frame component $\fc{R}^i\!_{jkl}$ diverges. This could 
happen if all sectional curvatures diverge, for example.

We are now able to show that the boundary fibres are totally 
degenerate in many cases.  We will employ the following procedure.  
Let $\gamma:I\to M$ be a curve with an endpoint $p\in\d M$, and let 
$E$ be a pseudo-orthonormal frame field on (a subset of) $M$.  Using 
Cartan's equations we find the rotation coefficients and the Riemann 
tensor components in the frame $E$.  We may then write down and solve 
the parallel propagation equations for a frame $F$ along $\gamma$.  
The tricky part is finding a sequence of parameter values $t_i$ along 
with suitable $\rho_i$-balls $\U_i$ and bounds on $\snormf{\U_i}{R}$ 
and $\snormf{\U_i}{\cov R}$.  To this end, we need to explore the 
connection between the b-distance and Lorentz transformations.

\begin{lemma}\label{la:Lbound}
Let $p \in M$ and $\V\subseteq B_\rho(p,E_p)\subset OM$, and suppose 
that $E_p$ can be extended to a frame field $E$ on $\V$. Put 
$\snormf{\pi(\V)}{\Gamma} := \sup_{\pi(\V)}\norm\Gamma$, where 
$\Gamma$ is the array of the rotation coefficients in the frame $E$,
and $K:=\max\{\snormf{\pi(\V)}{\Gamma},1\}$.  If $\rho \le 1/4K$ then 
all frames in $\V$ can be expressed as $E\fc{L}$ with $\normf{L}<2$.
\end{lemma}

\begin{proof}
Let $\kappa:[0,\rho]\to\V$ be a curve in $\V$, parameterised by 
b-length, with $\kappa(0)=(p,E)$.  Let $\dot\kappa$ be the tangent 
vector of $\kappa$, and let $V$ be the tangent vector of 
$\pi\circ\kappa$ with components $\fc{V}$ in the fixed frame $E$.  
Also, let the frame $F$ of $\kappa$ be given by 
$F=E\fc{L}$.  We want to show that $\norm{\fc{L}}<2$.

From \cite{Kobayashi-Nomizu-I}, the fundamental 1-form $\theta$ at 
$\kappa(s)$ is given by $F^{-1}\circ\pi_*$, where $F$ is regarded as a 
map $\R^4 \to T_{\pi\circ\kappa}M$, so
\begin{equation}\label{eq:thetakappa}
	\theta(\dot\kappa) = \fc{L}^{-1}\fc{V}.
\end{equation}
Next, the connection form $\omega$ is given by 
\begin{equation}\label{eq:connform}
	\varphi\bigl(\omega(\dot\kappa)\bigr) = \mbox{ver}\:\dot\kappa,
\end{equation}
where $\varphi$ is the canonical isomorphism from $\l$ to the vertical 
subspace of $T_{\kappa(s)}OM$, and $\mbox{ver}\:\dot\kappa$ denotes the 
vertical component of $\dot\kappa$ \cite{Kobayashi-Nomizu-I}.  By 
definition, if $a\in\l$ and $A(t)$ is any curve in $\L$ 
with $A(0)=\delta$ and $\frac{\id}{\id t}\bigr|_{t=0}A = a$, then
\begin{equation}\label{eq:canoniso}
	\varphi(a) := \frac{\id}{\id t}\biggr|_{t=0} R_{A(t)} F = F a
\end{equation}
at $F$. The vertical component of $\dot\kappa$ is given by
\begin{equation}\label{eq:dFds}
	\cov_{V}F = (\cov_{V}E)\fc{L} + E\dot\fc{L}	
	= F\fc{L}^{-1}( \Gamma\fc{V}\fc{L} + \dot{\fc{L}} ),
\end{equation}
where $\Gamma\fc{V}\fc{L}$ is the matrix with components 
$\Gamma^{i}\!_{kl}\fc{V}^k\fc{L}^l\!_j$ and $\Gamma^{i}\!_{kl}$ are the 
rotation coefficients of the frame $E$. Combining 
\eqref{eq:connform}, \eqref{eq:canoniso} and \eqref{eq:dFds} gives 
\begin{equation}\label{eq:omegakappa}
	\omega(\dot\kappa) = 
	\fc{L}^{-1} ( \Gamma\fc{V}\fc{L} + \dot{\fc{L}} ).
\end{equation}
Since $\kappa$ is parameterised by b-length, 
\begin{equation}
	\abs{\theta(\dot\kappa)}^2 + \norm{\omega(\dot\kappa)}^2 = 1,
\end{equation}
so from \eqref{eq:thetakappa}, 
\begin{equation}
	\absf{V} \le \normf{L}\abs{\theta(\dot\kappa)} \le \normf{L}
\end{equation}
and from \eqref{eq:omegakappa},
\begin{equation}\label{eq:normLde}
	\Abs{\frac{\id}{\id s}\normf{L}} 
	\le \normf{L}\norm{\omega(\dot\kappa)} + \norm\Gamma\absf{V}\normf{L}
	\le K\normf{L}^2 + \normf{L}.
\end{equation}
Put $u := K\normf{L}$. Then
\begin{equation}
	\frac{\dot{u}}{u^2+u} \le 1,
\end{equation}
and integration gives
\begin{equation}
	u \le \frac{K}{(K+1)e^{-s}-K}
\end{equation}
since $\normf{L}=1$ at $s=0$.
Thus 
\begin{equation}
	\normf{L} \le \bigl( (K+1)e^{-s}-K \bigr)^{-1},
\end{equation}
and the result follows from $s \le 1/4K$ and $K\ge1$.
\end{proof}

\begin{note}
\eqref{eq:normLde} corresponds to the 
differential equation on page~42 of \cite{Clarke:analysis-sing}, 
except that there the last term is $1$ instead of $\normf{L}$ which 
is incorrect.
\end{note}

\subsection{FLRW space-times}
\label{sec:FLRW}

Let $(M,g)$ be a Robertson-Walker space-time, i.e.\ $M=(0,\tau)\times\Sigma$ 
and $g$ is defined by the line element
\begin{equation}
	\id s^2=-\id t^2+a(t)^2\id\sigma^2
\end{equation}
such that $(\Sigma,\id\sigma^2)$ is a homogeneous space (see eg.\ 
\cite{Hawking-Ellis,Misner-Thorne-Wheeler,Clarke:analysis-sing}).  The 
scale function $a(t)$ is determined from the chosen matter model via 
the field equations.  For a Friedman big bang model, $a(t) \to 0$ as 
$t \to 0$, corresponding to a curvature singularity at $t=0$.

Let $\gamma$ be a curve in $M$ with constant projection $x\in\Sigma$, 
parameterised by $t$.  Then $\gamma$ starts at the singularity at 
$t=0$.  Choose the pseudo-orthonormal frame field $E$ on (a subset of) 
$M$ as
\begin{equation}
	E_0 := \frac\partial{\partial t}\qquad\text{and}\qquad
  E_\alpha := a(t)^{-1}\tilde{E}_\alpha
\end{equation}
where $\tilde{E}$ is an orthonormal frame field on the Riemannian 
manifold $(\Sigma,\id\sigma^2)$.  Note that $\tilde{E}$ may be defined 
only on a neighbourhood of $x$ if $(\Sigma,\id\sigma^2)$ does not 
admit a global parallelisation.  Here greek indices $\alpha, 
\beta,\dots$ refer to spatial components and have values in 
$\{1,2,3\}$.  Write $\theta$ for the cotangent frame field dual to 
$E$, i.e.\ $\theta$ is the fundamental 1-form restricted to the 
section of $OM$ defined by $E$.  From Cartan's equations, the 
nonvanishing connection and curvature form components are
\begin{equation}
\begin{aligned}
	\omega^0\!_\alpha &= \omega^\alpha\!_0 = 
		\dot aa^{-1}\,\theta^\alpha \\
	\omega^\alpha\!_\beta &= -\omega^\beta\!_\alpha = 
		a^{-1} \,\tilde\Gamma^\alpha\!_{\mu\beta}\,\theta^\mu
\end{aligned}
\end{equation}
and
\begin{equation}
\begin{aligned}
	\Omega^0\!_\alpha &= \Omega^\alpha\!_0 = 
		\ddot{a}a^{-1} \,\theta^0\!\wedge\theta^\alpha \\
	\Omega^\alpha\!_\beta &= -\Omega^\beta\!_\alpha = 
		(a^{-2} \,\tilde{\fc{R}}^\alpha\!_{\beta\mu\nu} + 
		\dot a^2a^{-2} \,\delta^\alpha\!_\mu\delta_{\beta\nu})
		\,\theta^\mu\!\wedge\theta^\nu
\end{aligned} 
\end{equation}
where a dot denotes differentiation with respect to $t$ and 
$\tilde{\Gamma}^\alpha\!_{\delta\beta}$ and 
$\tilde{\fc{R}}^\alpha\!_{\beta\mu\nu}$ are the rotation 
coefficients and the Riemann tensor components, respectively, of 
$(\Sigma,\id\sigma^2)$ in the frame $\tilde{E}$.

Solving the parallel propagation equations we find that $E$ is 
parallel along $\gamma$.  To study the asymptotic behaviour we 
consider the case $a(t)=t^c$ for a constant $c\in(0,1)$.  Then there 
are positive constants $N_1$ and $N_2$ such that
\begin{equation}
	\normf{R} < N_1 \max\{ t^{-2}, t^{-2c}\norm{\tilde{\fc{R}}} \}
\end{equation}
and
\begin{equation}
	\normf{\cov R} <
	N_2 \max\{ t^{-3}, t^{-2c-1}\norm{\tilde{\fc{R}}}, 
		t^{-3c}\norm{\cov\tilde{\fc{R}}} \}
\end{equation}
in the frame $E$. Moreover, $\fc{R}$ is invertible on $\gamma$ and 
\begin{equation}
	\norm{\fc{R}^{-1}} < N_3\,t^2
\end{equation}
for some positive constant $N_3$, so $\norm{\fc{R}^{-1}} \to 0$ as $t 
\to 0$.  Pick a sequence $t_i\to 0$ and let $\rho_i:=t_i/8$ and 
$\U_i:=B_{\rho_i}\bigl(\gamma(t_i)\bigr)$.  Let $\mathcal{S}$ be a 
neighbourhood of $x$ in $\Sigma$ such that $\norm{\tilde\Gamma}$, 
$\norm{\tilde{\fc{R}}}$ and $\norm{\cov\tilde{\fc{R}}}$ are bounded on 
$\mathcal{S}$.  Put $\V_i := \U_i\cap\K_i$, where 
$\K_i := \pi^{-1}([t_i/2,3t_i/2]\times\mathcal{S})$.  Then for small 
enough $t_i$,
\begin{equation}
	1 < \snormf{\V_i}{\Gamma} \le 2ct_i^{-1} = K_i
\end{equation}
and since $c<1$, $\rho_i<1/4K_i$.  Thus Lemma~\ref{la:Lbound} gives 
$\normf{L}<2$ on $\V_i$.  If $\kappa$ is a curve in $\V_i$ with 
$\kappa(0)=\gamma(t_i)$ and $l(\kappa)\le\rho_i$, the $t$-coordinate 
satisfies
\begin{equation}
	\abs{t-t_i} 
	= \Abs{\int_0^s \bigl(\fc{L}\theta(\dot\kappa)\bigr)^0 \,\id s}
	\le \normf{L} l(\kappa)
	< \frac{t_i}{4}
\end{equation}
on $\kappa$. Let $\tilde\kappa$ be the 
projection of $\pi\circ\kappa$ to $\Sigma$.  Since $\tilde{E}$ is an 
orthonormal frame, the metric length of $\tilde\kappa$ in 
$(\Sigma,\id\sigma^2)$ can be estimated by
\begin{equation}
	l_\sigma(\tilde\kappa) 
	\le \int_0^s a^{-1}\normf{L}\abs{\theta(\dot\kappa)} \,\id s
	< 2^{c-2}\,t_i^{1-c}
\end{equation}
which tends to 0 as $t_i \to 0$.  But then $\U_i$ must be contained 
in $\K_i$ for small enough $t_i$, so the $t$-coordinate 
must be greater than $t_i/2$ on the whole of $\U_i$.

Thus $\norm{\fc{R}_i^{-1}}^3 \snormf{\U_i}{R}^2$, 
$\norm{\fc{R}_i^{-1}}^2 \snormf{\U_i}{\cov R}$ and 
$\norm{\fc{R}_i^{-1}}/\rho_i$ all tend to 0 as $t_i \to 0$, so by 
Theorem~\ref{th:degeneracy} the fibre over $\gamma(0)$ is totally 
degenerate.  Note that in \cite{Clarke:analysis-sing}, a similar 
result is given for $2/3<c<1$. The reason for the restriction on $c$ 
is that Clarke uses a bound on $\norm{\fc{R}^{-1}}$ of order $t^{2c}$, 
while $\norm{\fc{R}^{-1}}$ is actually of order $t^2$ for small 
enough $t$.

\subsection{Kasner space-times}
\label{sec:Kasner}

To illustrate that the fibre degeneracy is not an artefact of isotropy 
we repeat the calculations for the anisotropic Kasner space-times (see 
e.g.\ \cite{Misner-Thorne-Wheeler}).  Let $M := I\times\Sigma$ with 
metric $g$ given by
\begin{equation}
	\id s^2=-\id t^2 + t^{2p_x}\id x^2 + t^{2p_y}\id y^2 + t^{2p_z}\id z^2,
\end{equation}
where $(x,y,z)$ are coordinates on $\Sigma$ and the constants $p_x$, 
$p_y$ and $p_z$ satisfy
\begin{equation}
	p_x+p_y+p_z=1\qquad\text{and}\qquad p_x^2+p_y^2+p_z^2=1.
\end{equation}
We exclude the special case when $p_x=p_y=0$, $p_z=1$ (including 
permutations of $x$, $y$ and $z$) which corresponds to one half of 
Minkowski space.  For all other parameter values, there is a curvature 
singularity at $t=0$.  Let $\gamma$ be a curve with constant $x$, $y$ 
and $z$, starting at the singularity and parameterised by $t$.  
Choosing a pseudo-orthonormal frame field $E$ as
\begin{equation}
	E_0 := \frac\d{\d t},\quad
  E_1 := t^{-p_x}\frac\d{\d x},\quad
  E_2 := t^{-p_y}\frac\d{\d y}\quad\text{and}\quad
  E_3 := t^{-p_z}\frac\d{\d z}
\end{equation}
we find again that $E$ is parallel propagated along $\gamma$, that 
$\fc{R}$ is invertible, and that $\normf{R} < N_1\,t^{-2}$, 
$\normf{\cov R} < N_2\,t^{-3}$ and $\norm{\fc{R}^{-1}} < N_3\,t^2$ for 
some constants $N_1$, $N_2$ and $N_3$.  Put 
$p:=\max\{\abs{p_x},\abs{p_y},\abs{p_z}\}$.  Then 
$\norm\Gamma=pt^{-1}$, and an argument similar to that in 
\S\ref{sec:FLRW} gives that the boundary fibre is totally degenerate.

\subsection{Schwarzschild space-time}
\label{sec:Schwarz}

Let $(M,g)$ be given by
\begin{equation}
	\id s^2=b(r)^{-2}\id t^2-b(r)^2\id r^2+
  r^2(\id\vartheta^2+\sin^2\!\vartheta\,\id\phi^2)
\end{equation}
with $t\in\mathbb{R}$, $r\in(0,2m)$, $\vartheta\in[0,\pi]$, 
$\phi\in[0,2\pi)$, and 
\begin{equation}
	b(r) := \Bigl( \frac{2m}{r} - 1 \Bigl)^{-1/2}
\end{equation}
(see e.g.~\cite{Hawking-Ellis,Misner-Thorne-Wheeler}).  Choose $E$ as
\begin{equation}
	E_0 := b^{-1}\frac\d{\d r},\quad
  E_1 := b\frac\d{\d t},\quad
  E_2 := r^{-1}\frac\d{\d\vartheta}\quad\text{and}\quad
  E_3 := (r\sin\vartheta)^{-1}\frac\d{\d\phi}
\end{equation}
and let the corresponding cotangent frame be $\theta$.  The connection 
form is
\begin{equation}
\begin{aligned}
	\omega^0\!_1 &= \omega^1\!_0 = -mbr^{-2} \,\theta^1 &\qquad
	\omega^0\!_2 &= \omega^2\!_0 = b^{-1}r^{-1} \,\theta^2 \\
	\omega^0\!_3 &= \omega^3\!_0 = b^{-1}r^{-1} \,\theta^3 &
	\omega^2\!_3 &= -\omega^3\!_2 = -r^{-1}\cot\vartheta \:\theta^3
\end{aligned}
\end{equation}
and the curvature form is
\begin{equation}
\begin{aligned}
	\Omega^0\!_1 &= \Omega^1\!_0 = 2mr^{-3} \,\theta^0\!\wedge\theta^1 &\qquad
	\Omega^0\!_2 &= \Omega^2\!_0 = -mr^{-3} \,\theta^0\!\wedge\theta^2 \\
	\Omega^0\!_3 &= \Omega^3\!_0 = -mr^{-3} \,\theta^0\!\wedge\theta^3 &
	\Omega^1\!_2 &= -\Omega^2\!_1 = -mr^{-3} \,\theta^1\!\wedge\theta^2 \\
	\Omega^1\!_3 &= -\Omega^3\!_1 = -mr^{-3} \,\theta^1\!\wedge\theta^3 &
	\Omega^2\!_3 &= -\Omega^3\!_2 = 2mr^{-3} \,\theta^2\!\wedge\theta^3.
\end{aligned}
\end{equation}
Thus there are positive constants $N_1$ and $N_2$ such that 
$\normf{R}<N_1\,r^{-3}$ and $\normf{\cov R}<N_2\,r^{-9/2}$ in the frame 
$E$.

Let $\gamma$ be a radial curve parameterised by $r$ with 
$\vartheta=\pi/2$, $\phi=0$ and $t=t_0$.  Then $E$ is parallel on 
$\gamma$, $\fc{R}$ is invertible, and $\norm{\fc{R}^{-1}}<r^3/m$ along 
$\gamma$.  If $\vartheta$ is bounded away from 0 and $\pi$,
\begin{equation}
	\norm{\Gamma} \le \sqrt{2m}r^{-3/2}
\end{equation}
for small $r$.  Choosing a sequence $r_i \to 0$ and 
\begin{equation}
	\rho_i := \frac{r_i^{3/2}}{16\sqrt{m}},
\end{equation}
an argument similar to that in \S\ref{sec:FLRW} gives 
$\normf{L}<2$ and $r>r_i/2$ on each 
$\U_i := B_{\rho_i}\bigl(\gamma(r_i)\bigr)$ for small enough $r_i$.  Thus 
the conditions of Theorem~\ref{th:degeneracy} are fulfilled, so the 
boundary fibre is totally degenerate.

\subsection{Reissner-Nordstr\"om space-time}
\label{sec:Reiss-N}

Let $(M,g)$ be given by
\begin{equation}
	\id s^2 = -b(r)^{-2}\id t^2 - b(r)^2\id r^2 + 
  r^2 ( \id\vartheta^2 + \sin^2\!\vartheta\,\id\phi^2 )
\end{equation}
with $t\in\mathbb{R}$, $r\in(0,r_-)$, $\vartheta\in[0,\pi]$ and 
$\phi\in[0,2\pi)$, and 
\begin{equation}
	b(r) := \biggl( 1 - \frac{2m}{r} + \frac{e^2}{r^2} \biggr)^{-1/2}
\end{equation}
(see e.g.\ \cite{Hawking-Ellis,Misner-Thorne-Wheeler}).  Degeneracy of 
the boundary fibre follows directly by generalising the argument in 
\S\ref{sec:Schwarz}, with $\rho_i:=r_i^2/32\abs{e}$, 
$\normf{R}<N_1\,r^{-4}$, $\normf{\cov R}<N_2\,r^{-6}$ and 
$\norm{\fc{R}^{-1}}<N_3\,r^4$.  Note that the timelike nature of the 
singularity does not affect the argument.

\subsection{Tolman-Bondi space-time}
\label{sec:Tol-Bondi}

The metric for the spherically symmetric Tolman-Bondi space-time 
$(M,g)$ is given by 
\begin{equation}
	\id s^2 = -\id t^2 + e^{2\omega}\id r^2 + 
	R^2(\id\vartheta^2 + \sin^2\!\vartheta\,\id\phi^2)
\end{equation}
where $\omega:=\omega(t,r)$ and $R:=R(t,r)>0$ 
\cite{Newman:Tolman-Bondi}.  If the energy momentum tensor is taken to 
be of dust form,
\begin{equation}
	T := \epsilon(t,r) \,\frac\d{\d t}\otimes\frac\d{\d t},
\end{equation}
the equations for $\omega$ and $R$ are
\begin{gather}
	\frac12 \dot{R}^2 - \frac{m}{R} = \frac12 (W^2-1) \label{eq:Tol-Bondi} \\
	R' = We^\omega \\
	\epsilon = \frac{r^2 \rho}{R^2 R'}
\end{gather}
where $W:=W(r)$, $\rho(r):=\epsilon(0,r)$, dots and primes denote 
partial derivatives with respect to $t$ and $r$ respectively, and
\begin{equation}
	m(r) := 4\pi \!\int_0^r\!\! \rho r^2 \,\id r.
\end{equation}
Here $r$ is rescaled such that $r:=R(0,r)$ and $\omega$, $R$ and 
$\epsilon$ are assumed to be smooth functions of $t$ and $r$.  For 
physical reasons, we require that $\epsilon(t,r)\ge0$ and 
$\epsilon(t,0)>0$.  Put
\begin{equation}
	E(r) := \frac12\bigl(W^2(r)-1\bigr)
\end{equation}
and let
\begin{equation}
	a(r) := \frac{3m(r)}{4\pi r^3}
\end{equation}
and
\begin{equation}
	p(r) := -\frac{E(r)R(0,r)}{m(r)}.
\end{equation}
It can be shown that both $a$ and $p$ extend to smooth even functions 
of $r$ on $\R$, with $a(r) > 0$ and $p(r) \le 1$.

Choose a pseudo-orthonormal frame $E$ with cotangent frame $\theta$ 
according to
\begin{equation}
	E_0 := \frac{\d}{\d t},\quad
  E_1 := \frac{W}{R'}\frac{\d}{\d r},\quad
  E_2 := R^{-1}\frac{\d}{\d\vartheta},\quad\text{and}\quad
  E_3 := (R\sin\vartheta)^{-1}\frac{\d}{\d\phi}.
\end{equation}
Then the connection form is 
\begin{equation}
\begin{aligned}
	\omega^0\!_1 &= \omega^1\!_0 = \frac{\dot{R}'}{R'} \,\theta^1 &\qquad
	\omega^0\!_2 &= \omega^2\!_0 = \frac{\dot{R}}{R} \,\theta^2 &\qquad
	\omega^0\!_3 &= \omega^3\!_0 = \frac{\dot{R}}{R} \,\theta^3 \\
	\omega^1\!_2 &= -\omega^2\!_1 = -\frac{W}{R} \,\theta^2 &
	\omega^1\!_3 &= -\omega^3\!_1 = -\frac{W}{R} \,\theta^3 &
	\omega^2\!_3 &= -\omega^3\!_2 = -\frac{1}{R}\cot\vartheta \:\theta^3
\end{aligned}
\end{equation}
and the curvature form is
\begin{equation}
\begin{aligned}
	\Omega^0\!_1 &= \Omega^1\!_0 = 2mR^{-3} \,\theta^0\!\wedge\theta^1 &\qquad
	\Omega^0\!_2 &= \Omega^2\!_0 = -mR^{-3} \,\theta^0\!\wedge\theta^2 \\
	\Omega^0\!_3 &= \Omega^3\!_0 = -mR^{-3} \,\theta^0\!\wedge\theta^3 &
	\Omega^1\!_2 &= -\Omega^2\!_1 = 
		\Bigl(\frac{m'}{R'R^2}-\frac{m}{R^3}\Bigr) \,\theta^1\!\wedge\theta^2 \\
	\Omega^1\!_3 &= -\Omega^3\!_1 = 
		\Bigl(\frac{m'}{R'R^2}-\frac{m}{R^3}\Bigr) \,\theta^1\!\wedge\theta^3 &
	\Omega^2\!_3 &= -\Omega^3\!_2 = 2mR^{-3} \,\theta^2\!\wedge\theta^3.
\end{aligned}
\end{equation}
Integrating \eqref{eq:Tol-Bondi}, we get the following implicit 
expression for $R$.
\begin{equation}\label{eq:Rdef}
	\Bigl(\frac{R}{r}\Bigr)^{3/2} F(pR/r) = 
	F(p) - \frac{t}{t_0}\Bigl(\frac{a}{a_0}\Bigr)^{1/2} F(p_0)
\end{equation}
where $a_0:=a(0)=\rho(0)>0$, $p_0:=p(0)\le1$, $t_0:=(3/8\pi 
a_0)^{1/2}F(p_0)$, and $F:(-\infty,1)\to(0,\pi/2)$ is a positive, 
bounded, smooth, strictly increasing and strictly convex function.  If 
$E(r)<0$, \eqref{eq:Rdef} is singular on a hypersurface $\{t=t_b(r)\}$ 
where $pR=r$, with $t_b(r)\le0$.  For $t<t_b(r)$ an equation similar 
to \eqref{eq:Rdef} holds, and we will concentrate on the region where 
$t>0$.  We refer to \cite{Newman:Tolman-Bondi} for the details.

There are several types of singularities in the Tolman-Bondi 
space-time.  There is a coordinate singularity at $r=0$, a central 
singularity at $(t,r)=(t_0,0)$, and a final singularity at $r>0$, 
$R=0$.  For some parameter values, there are also shell crossing 
singularities where $R'=0$ (see \S\ref{sec:shell} below).

First we study the final singularity.  Let $\gamma$ be a curve with 
constant $r$, $\vartheta$ and $\phi$, and parameterise $\gamma$ by 
$\tau := t_s - t$, where
\begin{equation}
	t_s := \Bigl(\frac{a}{a_0}\Bigr)^{1/2} \frac{F(p)}{F(p_0)} \,t_0.
\end{equation}
Then $\gamma$ starts at the final singularity at $\tau=0$ and $E$ is 
parallel along $\gamma$.  All functions not depending on $t$ are 
bounded, so from \eqref{eq:Rdef} and \eqref{eq:Tol-Bondi}, there are 
constants $N_1$, $N_2$ and $N_3$ such that
\begin{align}
	\normf{R} &< N_1 \,\tau^{-2} \\
	\normf{\cov R} &< N_2 \,\tau^{-3} \\
	\norm{\fc{R}^{-1}} &< N_3 \,\tau^2.
\end{align}
By an argument as in \S\ref{sec:FLRW}, fibres over the final 
singularity are degenerate.

Next we turn our attention to the central singularity at $(t_0,0)$.  
Let $\gamma$ be a radial curve with $t=t_0$ and constant $\vartheta$ 
and $\phi$, parameterised by $r$ and starting at $(t,r)=(t_0,0)$.  
Also, let $F=E\fc{L}$ be parallel along $\gamma$.  Solving the parallel 
propagation equation we find that $\fc{L}$ is a Lorentz boost in the 
$(E_0,E_1)$-plane with hyperbolic angle
\begin{equation}
	\varphi := -\!\int \frac{\dot{R}'}{W} \,\id r.
\end{equation}
Let 
\begin{equation}
	C_0 := \frac12 p''(0) F'(p_0) - \frac{1}{4a_0} a''(0) F(p_0).
\end{equation}
We assume that $C_0 \neq 0$, the case of interest being $C_0>0$ since 
then the singularity is naked \cite{Newman:Tolman-Bondi}.  If we 
restrict attention to the neighbourhood of $\gamma$ where
\begin{equation}
	\abs{t-t_0} < \frac{C_0 \,t_0}{3F(p_0)}r^2,
\end{equation}
then it is possible to use \eqref{eq:Rdef}, \eqref{eq:Tol-Bondi} and 
the fact that $a$ and $p$ extends to $\R$ to estimate all components of 
$\Gamma$, $\fc{R}$ and $\cov\fc{R}$. We find that there are 
positive constants $N_1$, $N_2$ and $N_3$ such that 
\begin{align}
	\normf{R} &< N_1 \,r^{-4} \\
	\normf{\cov R} &< N_2 \,r^{-19/3} \\
	\norm{\fc{R}^{-1}} &< N_3 \,r^4.
\end{align}
Also, $\varphi$ is bounded as $r\to0$.  Again, an argument similar to 
the one in \S\ref{sec:FLRW}, with $\rho_i$ proportional to 
$r_i^{7/3}$, gives that the fibre is totally degenerate also for this 
naked singularity.  Note that $\normf{\cov R}$ has a stronger 
divergence than $\normf{R}^{3/2}$.

\section{Partial degeneracy}
\label{sec:part-degen}

In general it can be very hard to show that a boundary fibre is 
degenerate, since different subgroups of the singular holonomy group 
may be generated by various things, e.g.\ unbounded curvature, regular 
curvature, quasi-regular singularities, and contributions from 
other boundary points due to non-Hausdorff behaviour of the b-boundary 
\cite{Clarke:sing-holonomy}.  Note that even if the Riemann tensor is 
non-invertible and/or if only some components diverge, in some cases 
Lemma~\ref{la:gener1} may be used to establish partial degeneracy at 
least.

\subsection{Quasi-regular singularities}
\label{sec:quasi}

To illustrate how degeneracy can be caused by topological anomalies we 
consider quasi-regular singularities obtained by suitable 
identifications in (the universal covering space of) Minkowski 
space-time $(M,g)$ with 
\begin{equation}
	\id s^2 = - \id t^2 + \id x^2 + \id y^2 + \id z^2.
\end{equation}
Given an isometry $\varphi$, we may identify points $\varphi(p)$ 
with $p$ in (the universal covering space of) a subset of $(M,g)$ 
\cite{Ellis-Schmidt:singular}.

As a first example, let $(\hat{M},\hat{g})$ be the universal covering 
space of Minkowski space with the timelike 2-plane $\{x=y=0\}$ removed 
and let $\varphi$ be the rotation in the $(x,y)$-plane by an angle 
$\phi \ne2 \pi$.  Then the space-time obtained by identifying points 
with their images under $\varphi$ has a conelike singularity at 
$\{x=y=0\}$.  Since $(\hat{M},\hat{g})$ is flat, the infinitesimal, 
local and restricted holonomy groups are all trivial, so the only 
contribution to the singular holonomy group comes from curves not 
homotopic to $0$.  It clearly suffices to study curves with 
$x^2+y^2=r$ as $r\to0$, and a simple argument then gives that 
$\Phis{OM}$ is a discrete group generated by $\phi$ modulo $2\pi$.

Secondly, let $\varphi$ be a boost in the $(t,x)$-plane with 
hyperbolic angle $\phi$ and consider the subset $\{z>-t\}$ of $(M,g)$.  
Identifying points under $\varphi$ we get the Misner space-time with 
quasi-regular singularities similar to the ones in the Taub-NUT 
space-time \cite{Hawking-Ellis,Misner:Taub-NUT}.  As for the conelike 
example above, it is straightforward to show that $\Phis{OM}$ is 
generated by $\varphi$.  More complicated singularities can be 
constructed by variations of this procedure 
\cite{Ellis-Schmidt:singular}.

\subsection{Shell crossing singularities}
\label{sec:shell}

We return to the Tolman-Bondi space-time from 
\S\ref{sec:Tol-Bondi} to study the shell crossing singularities 
where $R'=0$.  Only some components of the curvature diverge, so all 
we can hope for is to establish partial degeneracy in some directions.  
Unfortunately, it turns out that while $\normf{R}$ is of order 
$(R')^{-1}$, $\normf{\cov R}$ is of order $(R')^{-3}$, which prohibits 
us from using Lemma~\ref{la:gener1} in this case.  Also, higher order 
derivatives of the Riemann tensor have even stronger divergence.  
Since the infinitesimal holonomy group is generated by the Riemann 
tensor and its derivatives, whose norms all diverge, it seems probable 
that the singular holonomy group is nontrivial.  Proving that is 
impossible with our technique however, since we have no way to control 
the contributions from higher order terms.

\section{Discussion}
\label{sec:disc}

We have shown that in many cases, the b-boundary have totally 
degenerate fibres, leading to undesired topological effects.  The 
argument is based on that the divergence of the derivative of the 
Riemann tensor is sufficiently weak, so that the essential 
contribution to the singular holonomy group comes from 
$\fc{R}(\fc{X},\fc{Y})$.  As we saw in \S\ref{sec:shell}, this 
fails in some cases.  Since the infinitesimal holonomy group is 
generated by expressions of the form 
$\cov_{\fc{V}_1\dots\fc{V}_n}(\fc{X},\fc{Y})$, it might be possible to 
use higher order derivatives of the Riemann tensor to generate 
elements in the singular holonomy group.  One would then have to go 
further in the expansion in the proof of Lemma~\ref{la:approx}, and 
the conditions would get much more complicated.

In \S\ref{sec:quasi}, we gave a simple example of how a 
quasi-regular singularity can give rise to degenerate fibres.  It is 
very easy to construct examples of quasi-regular singularities with 
discrete singular holonomy groups, but it is unknown if nondiscrete 
groups can arise in this way.

The most apparent unsolved problem involving the b-boundary is the 
structure of the boundary itself.  In the FLRW case the boundary has 
been shown to be a single point \cite{Clarke:analysis-sing}.  But for 
the Schwarzschild space-time, the results are not as conclusive.  Both 
Bosshard \cite{Bosshard:b-boundary} and Johnson 
\cite{Johnson:b-boundary} have established partial degeneracy of 
boundary fibres, but it is unknown if the boundary is just a point or 
something else (Johnson conjectures that it is a line).

\end{document}